\documentclass[conference]{IEEEtran}
\IEEEoverridecommandlockouts
\usepackage{amsmath,amsfonts,amssymb, amsthm}
\usepackage{array}
\usepackage{booktabs}
\usepackage{textcomp}
\usepackage{stfloats}
\usepackage{url}
\usepackage{verbatim}
\usepackage{balance}  
\usepackage{graphicx}
\usepackage{cite}
\usepackage{subcaption}
\usepackage{algorithm}
\usepackage{adjustbox}
\usepackage{algpseudocode}
\usepackage{caption}
\usepackage[normalem]{ulem} 
\usepackage[bookmarks=false]{hyperref}
\hypersetup{
	colorlinks = true,
	linkcolor = blue,
	urlcolor =  blue,
	citecolor = blue
}

\pdfoutput=1

\begin{document}
\captionsetup[figure]{font={small}, name={Fig.}, labelsep=period}

\title{Beamforming-Codebook-Aware Channel Knowledge Map Construction for Multi-Antenna Systems}

	\author{
		Haohan~Wang\textsuperscript{1},
		~Xu~Shi\textsuperscript{1},~Hengyu~Zhang\textsuperscript{1},~Yashuai~Cao\textsuperscript{1},~Jintao~Wang\textsuperscript{1}\\
		\IEEEauthorblockA{
		\textsuperscript{1}Beijing National Research Center for Information Science and Technology (BNRist),\\
		Dept. of Electronic Engineering, Tsinghua University, Beijing, China\\
		\{whh24@mails., shi-x@mail., zhanghen23@mails., caoys@, wangjintao@\}tsinghua.edu.cn}}

\maketitle

\begin{abstract}
Channel knowledge map (CKM) has emerged as a crucial technology for next-generation communication, enabling the construction of high-fidelity mappings between spatial environments and channel parameters via electromagnetic information analysis. Traditional CKM construction methods like ray tracing are computationally intensive. Recent studies utilizing neural networks (NNs) have achieved efficient CKM generation with reduced computational complexity and real-time processing capabilities. Nevertheless, existing research predominantly focuses on single-antenna systems, failing to address the beamforming requirements inherent to MIMO configurations. Given that appropriate precoding vector selection in MIMO systems can substantially enhance user communication rates, this paper presents a TransUNet-based framework for constructing CKM, which effectively incorporates discrete Fourier transform (DFT) precoding vectors. The proposed architecture combines a UNet backbone for multiscale feature extraction with a Transformer module to capture global dependencies among encoded linear vectors. Experimental results demonstrate that the proposed method outperforms state-of-the-art (SOTA) deep learning (DL) approaches, yielding a 17\% improvement in RMSE compared to RadioWNet. The code is publicly accessible at https://github.com/github-whh/TransUNet.
\end{abstract}


\section{Introduction}
The acquisition of a comprehensive and high-fidelity channel knowledge map (CKM) constitutes a pivotal enabler for the sixth-generation (6G) communication systems, facilitating spectrum sensing, beam management, and so on\cite{CKM}. Via the storage of environmental prior information, CKM provides essential channel state information (CSI) for pilot reduction and sumrate enhancement. Conventional channel modeling and mapping methods predominantly rely on ray-tracing techniques, whereas substantial computational requirements and inherent latency render them infeasible for dynamic communication environments \cite{Raytracing}. Recently, channel twinning paradigm \cite{Magazine} has emerged as a promising alternative, wherein CKM construction can be implemented effectively by environment-aware exploitation via deep learning.

Currently, CKM mapping approaches are predominantly categorized into two classes. The first encompasses pathloss maps, which are confined to single-antenna systems and thus fail to characterize spatial degrees of fredom such as beamforming parameters \cite{RadioUnet,GAN,WIFI}. In fact, the current neural-network-based CKM approaches lack careful architecture design for the specific beam pattern and propagation mechanism. The second category comprises various channel parameters, which are intensively dependent on ray-tracing techniques for construction. The concept of channel angle map (CAM) and beam index map (BIM) is propsoed \cite{CAM} without consideration of construction procedure. CKMImageNet \cite{Dataset} provides one database including angle-of-arrival (AoA) and angle-of-departure (AoD) only through ineffective ray-tracing approach. 
Most crucially, it is chaotic and computation-intensive to simultaneously characterize path-specific parameters with entirely distinct eigencharacteristics. These disadvantages limit the extension to CKM construction of multi-antenna systems.

The construction of CKMs for multi-antenna systems is promising but with critical challenges. As illustrated in Fig. 1, a beam-aware CKM utilizing a discrete Fourier transform (DFT) precoding vector exhibits significant discrepancies compared to the beam-unaware counterpart generated with a uniform precoding vector, underscoring the necessity of beamforming integration in the CKM design. Nevertheless, multi-antenna CKM construction suffers from the following challenges: 1) phase ambiguity due to the map resolution and positioning error, which deteriorates the multi-path coherent beamforming; 2) the construction inconsistency for several parameters with entirely distinct eigencharacteristics, such as delays, angles, and pahtloss; 3)  dimension mismatch and storage disorder for multiple paths among geographical sampling points, which is absolutely critical in practical communciations. To the best of our knowledge, no prior work has addressed beamforming-aware CKM  construction for massive MIMO systems.

\begin{figure}[!t]
	\centering
	\includegraphics[scale=0.38]{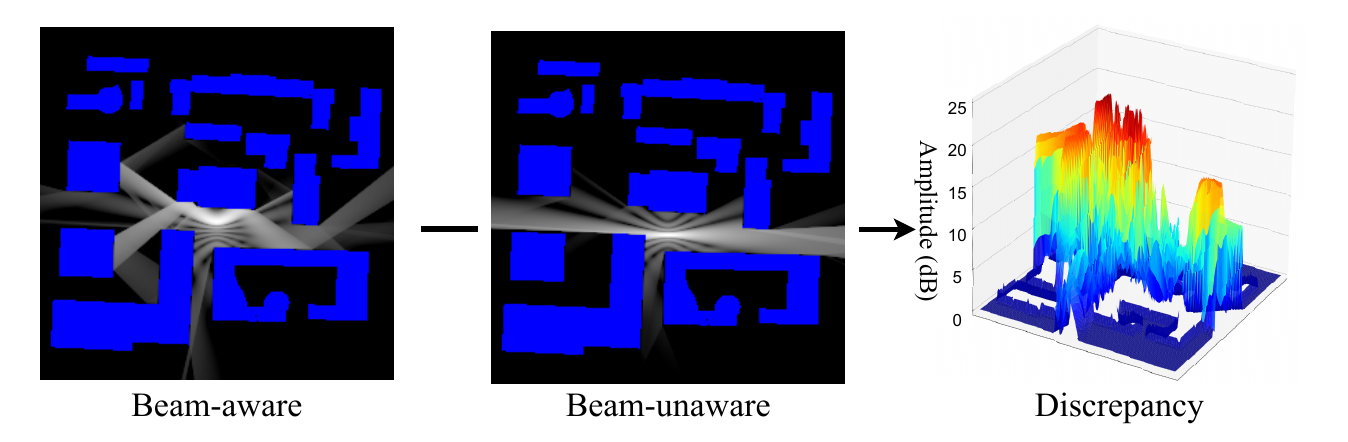} 
	\caption{Comparison of beam-aware and beam-unaware pathloss.}
	\label{fig1}
\end{figure}

Consequently, this paper endeavors to tackle two principal challenges in CKM construction. Primarily, conventional CKMs exhibit limited capability in precisely characterizing phase parameters, constrained by inherent deficiencies in mapping resolution and localization accuracy. This fundamental limitation consequently restricts measurable outputs to non-coherent physical parameters, notably pathloss and angular characteristics. Secondarily, given the practical implementation of finite static beamforming codebooks in realistic MIMO architectures, we develop a codebook-specific methodology that demonstrates both practical applicability and operational efficacy. This approach facilitates the reduction of equivalent channel representation to pathloss characterization across individual configurations.
The principal contributions are as follows:

\begin{itemize}

\item We present a beamforming-codebook-aware CKM construction framework by reformulating the traditional one from $\mathcal{F}(\mathbf{p})$ into an enhanced beam-dependent representation $\mathcal{F}(\mathbf{p}; \mathbf{w})$. This framework intrinsically incorporates beamforming constraints while eliminating reliance on imprecise and non-robust measurement information, thereby enabling accurate and implementation-feasible CKM construction that maintains full compatibility with practical beamforming implementations. 

\item We develop an environment-attention-based mapping approach via neural network named TransUNet, wherein the beam propagation characteristic is exploited. The architecture synergistically combines feature extraction, hierarchical spatial reconstruction, and multi-resolution interaction of environmental layouts 
from CNN, UNet and Transformer, respectively. Besides, we design one enhanced loss function integrating Sobel gradient operators with Laplacian pyramid decomposition to simultaneously optimize edge effects and beamforming-specific propagation pattern characterization. 

\item We establish an open-source dataset via ray-tracing-based Sionna simulations that systematically incorporates diverse precoding vectors. This comprehensive dataset accurately models beam propagation characteristics under varying electromagnetic (EM) material properties and architectural distributions, comprising 10,000 distinct scenarios that capture multi-beam spatial propagation patterns across heterogeneous environments.

\end{itemize}

The rest of the paper is organized as follows. Section II presents the system model. Section III demonstrates the beamforming codebook-specific CKM framework. Section IV elaborates on the TransUNet model and loss function. Section V provides the simulation results, followed by the conclusions in Section VI.

\begin{figure}[!t]
	\centering
	\includegraphics[scale=0.3]{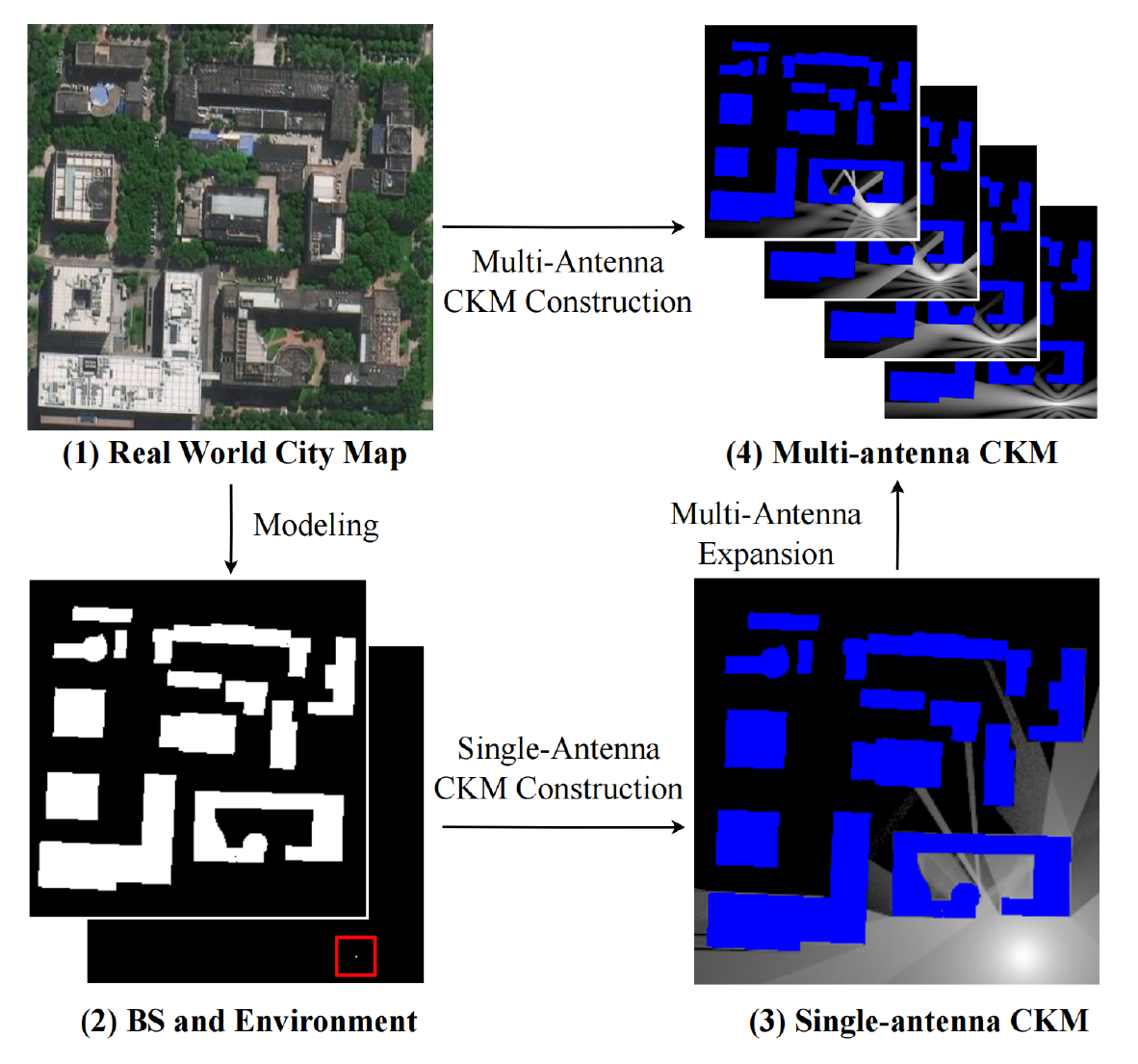} 
	\caption{Schematic diagram of real-world modeling and CKM construction based on the TransUNet.}
	\label{fig2}
\end{figure}

\section{System Model}

We consider a downlink MISO system where the BS is equipped with $N$ antennas arranged in a uniform linear array (ULA) and serves multiple single-antenna users. The transmitted symbol for user $i$ is expressed as $s_i$, so the data sent from BS can be written as $\mathbf{x}_i = \mathbf{w}_is_i$, where $\mathbf{w}_i$ is the precoding vector for the $i$-th user with $||\mathbf{w}_i||_2 = 1$. Supposing the channel vector between the BS and the $i$-th user at position $\mathbf{p}_i = (x_i, y_i, z_i)$ is expressed as $\mathbf{h}_i \in \mathbb{C}^{N \times 1}$, the received signal is given by

\begin{equation}
	y_i = \mathbf{h}^H_i\mathbf{w}_i s_i + z_i,
\end{equation}
where $z_i \sim \mathcal{CN}(0, \sigma^2)$ denotes additive white Gaussian noise (AWGN). The channel $\mathbf{h}_i$ can be further expressed as
\begin{equation}
	\mathbf{h}_i = \mathbf{h}_i^{\mathrm{LoS}} + \sum_{k=1}^{K} \mathbf{h}_{i,k}^{\mathrm{NLoS}},
\end{equation}
where $\mathbf{h}_i^{\mathrm{LoS}}$ denotes the \emph{line-of-sight} (LoS) component corresponding to direct propagation, $\mathbf{h}_{i,k}^{\mathrm{NLoS}}$ represents the $k$-th \emph{non-line-of-sight} (NLoS) component arising from electromagnetic wave reflections and scattering \cite{38901}. While this model provides detailed channel characterization, it is limited due to the large amount of parameters. To address this limitation, we propose a refined pathloss construction methodology for CKMs that captures spatial variations.


For the CKM formulation, we denote the spatial convergence of the region of interest by $\mathcal{M}$. The sampling process employs a uniform rectangular grid with horizontal and vertical sampling intervals of $d_x$ and $d_y$, yielding $N_x = W/d_x$ and $N_y = H/d_y$ sampling points along each dimension. All sampling is conducted at a fixed altitude. Each realization of $\mathcal{M}$ explicitly incorporates both the BS locations and building distributions. The construction procedures for both single-antenna and multi-antenna CKMs with equivalent channel gain parameters are illustrated in Fig. 2.


\section{Beamforming Codebook-Specific CKM}

In MIMO systems, the design and implementation of precoding vectors are of fundamental importance for achieving optimal spectral efficiency and mitigating interference \cite{Beamforming}. In this paper, the DFT beamforming vector can be expressed as
\begin{equation}
	\small
	\left.\mathbf{W}=\frac{1}{\sqrt{N}}\left(
	\begin{array}
		{ccccc}1  & \cdots & 1 \\
		\mathrm{e}^{j\pi\sin(\phi_0)}  & \cdots & \mathrm{e}^{j\pi\sin(\phi_{N-1})} \\
		\mathrm{e}^{j2\pi\sin(\phi_0)}  & \cdots & \mathrm{e}^{j\pi2\sin(\phi_{N-1})} \\
		\vdots & \vdots & \vdots \\
		\mathrm{e}^{j(N-1)\pi\sin(\phi_0)} & \cdots & \mathrm{e}^{j(N-1)\pi\sin(\phi_{N-1})}
	\end{array}\right.\right),
\end{equation}
where $\phi_i, \forall i$ is the angle of departure of the $i$-th BS antenna.

The primary objective of beamforming-aware CKM construction is to achieve precise pathloss estimation at each spatial coordinate $\mathbf{p}$ conditioned on a given precoding vector configuration. For an arbitrary $\mathbf{p}_i \in \mathcal{M}$, we aim to predict the pathloss through the NN $\mathcal{F}_{\Theta}$, where $\Theta$ parameterizes the network architecture. The complete beamforming-specific CKM is generated by evaluating $\mathcal{F}_{\Theta}$ over the entire spatial domain $\mathcal{M}$. In the context of single-antenna systems, this estimation process admits the following mathematical representation
\begin{equation}
	\mathbf{\mathrm{CKM}} = \mathcal{F}_{\Theta}(\mathbf{p}_i \in \mathcal{M}), \forall i.
\end{equation}
However, if the precoding vectors are considered, the CKM construction problem is given by
\begin{equation}
	\mathbf{\mathrm{CKM}}(j) = \mathcal{F}_{\Theta}(\mathbf{p}_i \in \mathcal{M}; \mathbf{w} = \mathbf{W}[:,j]), \forall i, j,
\end{equation}
where $j$ denotes the beamforming codeword index.

This task presents substantial technical challenges due to building-induced multipath reflections and scattering effects that distort electromagnetic wave propagation characteristics, consequently introducing considerable complexity in spatial pathloss estimation. Our objective is formulated as the minimization of the construction error between estimated and ground-truth pathloss values through the proposed TransUNet architecture, which can be written as
\begin{equation}
	\min_\Theta|\mathbf{\mathrm{CKM}}_{T}(j)-\mathcal{F}_\Theta(\mathbf{p}_i \in \mathcal{M}; \mathbf{w} = \mathbf{W}[:,j])|, \forall i, j, 
\end{equation}
where $\mathbf{\mathrm{CKM}}_{T}(j)$ is the ground truth of CKM  with the $j$-th precoding vector.

The construction of beam-specific CKMs offers transformative operational advantages in MIMO systems by enabling beam management, which is promising and instructive for the communication modules such as beam management, channel estimation and spectrum sensing. The technical superiority can be summarized as follows:
\begin{itemize}
	\item \textbf{Unified CKM parameter structure}: It enables structured encoding of channel parameters (e.g., angle, delay, path loss) via discrete beam codebooks. The pathloss-based structure is with unified dimension and similar eigencharacteristics, which is convenient for the batch training, inference and prediction via model/data-dirven approach. Thus the storage consumption and computational overhead are also reduced.
	
	\item \textbf{Incoherent pathloss expression}: It facilitates phase-insensitive path loss estimation while involves the angular beam information. Due to the limitation of CKM resolution and positioning accuracy, it is infeasible and hugely-costly to construct one phase-coherent CKM. In another word, the robustness and accuracy are significantly improved through this beam-specific CKM style.
	
	\item \textbf{Finite codebook-oriented implementation}: It is consistent to the real-world industry deployments with finite beam codebook, including mmWave and THz communication scenarios. The sum-rate and SINR metrics are available under this framework, while the BIM and CAM fail due to the path-specific mapping without phase alignment. Besides, it supports low-overhead CKM construction and update in dynamic environments (e.g., mobile/users) by leveraging a few codewords.
\end{itemize}

\section{TransUNet-based CKM construction}
In this section we develop an environment-attention-based mapping approach via neural network named TransUNet, wherein the beam propagation characteristic is exploited. The TransUNet first employs CNNs to extract building image features, then utilizes the Transformer layer with multihead self-attention (MHSA) and multi-layer perceptron (MLP) modules to model cross-region dependencies and finally adopts UNet's upsampling structure as the decoder. The details are demonstrated in the Fig. 3.

\begin{figure*}[!t]
	\centering
	\includegraphics[scale=0.30]{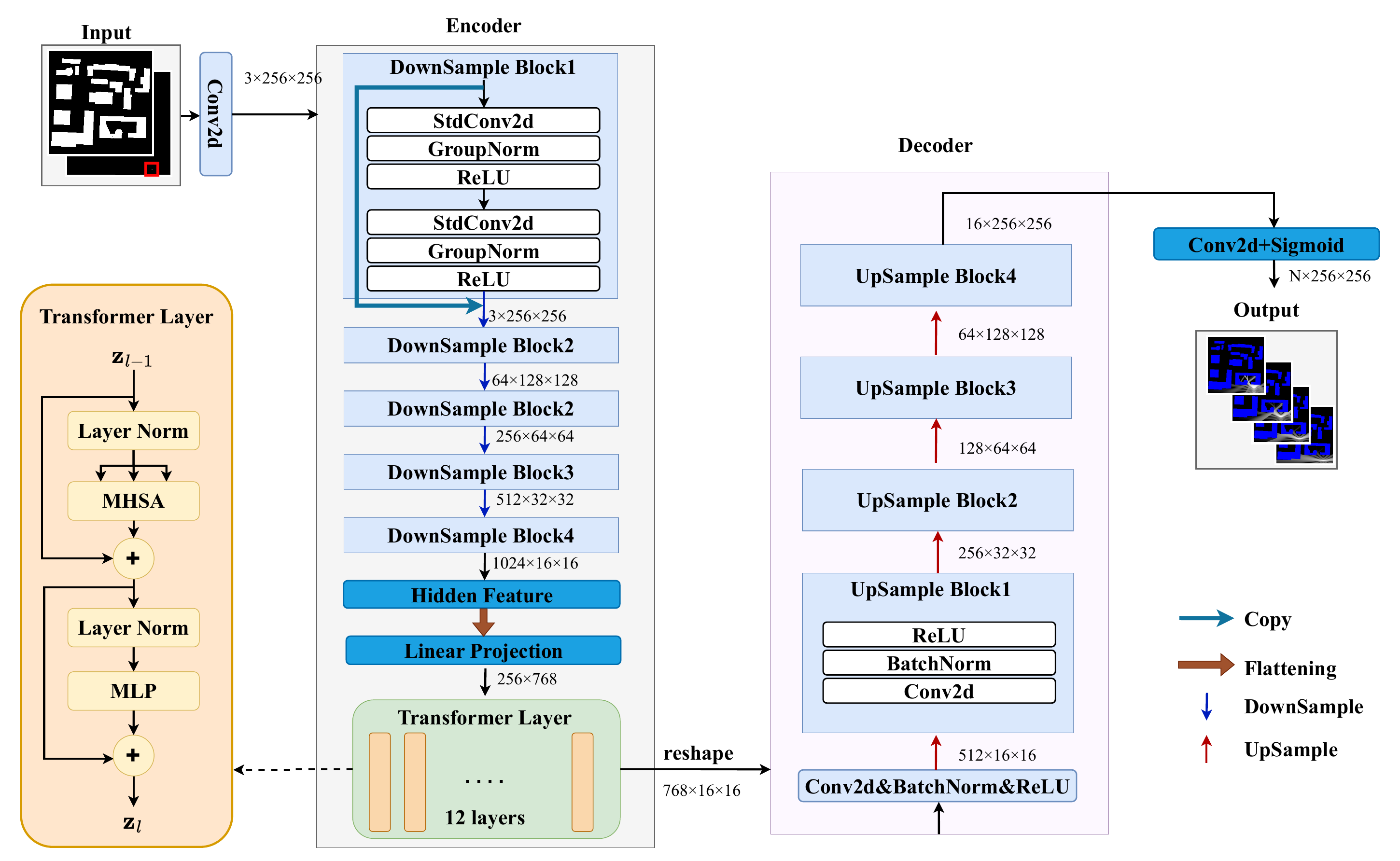} 
	\caption{The detailed structural schematic of TransUNet for CKM construction: CNN extracts hidden features, the Transformer layer learns contextual relationships, and the upsampling module reconstructs the CKM.}
	\label{fig3}
\end{figure*}

\subsection{CNN-Transformer Hybrid as Encoder}
The architectural implementation of TransUNet shares fundamental similarities with standard Vision Transformers (ViTs) in its utilization of patch embedding, MHSA, and MLP modules. To establish a technical foundation, we first delineate the canonical ViT framework before highlighting the distinctive characteristics of TransUNet. 

The fundamental procedure of ViT is as follows. Considering the $C \times H \times W$-dimensional input, the image is first partitioned into $P \times P$ patches, converting the $C \times H \times W$ input into $C \times N \times P \times P$ patch blocks, where $C$ is the input channel and $N = \frac{HW}{P^2}$. Then, these blocks are flattened to $\mathbf{x}_p^1, \mathbf{x}_p^2, ..., \mathbf{x}_p^N$, where $\mathbf{x}_p^i \in \mathbb{R}^{(P^2\cdot C)}, \forall i$. Then patches $\mathbf{x}_p^i, \forall i$ undergo the patch embedding module, which is given by

\begin{equation}
	\mathbf{z}_0=[\mathbf{x}_p^1\mathbf{E};\mathbf{~x}_p^2\mathbf{E};\cdots;\mathbf{~x}_p^N\mathbf{E}]+\mathbf{E}_{\text{pos}},
\end{equation}
where $\mathbf{E}\in\mathbb{R}^{(P^2\cdot C)\times D}$ denotes the patch embedding projection, and $\mathbf{E}_{\text{pos}}\in\mathbb{R}^{N\times D}$ denotes the position embedding, which is used to maintain the positional information.
The embedded sequence $\mathbf{z}_0$ subsequently propagates through Transformer modules that capture contextual relationships via attention mechanisms. The process of the $l$-th layer can be written as follows
\begin{equation}
	\begin{aligned}
		& \mathbf{z}_l^{\prime}=\mathrm{MHSA}(\mathrm{LN}(\mathbf{z}_{l-1}))+\mathbf{z}_{l-1}, \\
		& \mathbf{z}_l=\mathrm{MLP}(\mathrm{LN}(\mathbf{z}_l^{\prime}))+\mathbf{z}_l^{\prime},
	\end{aligned}
\end{equation}
where $\mathrm{LN}$ denotes the layer normalization. The structure of a Transformer layer is demonstrated in the left panel of Fig. 3.

In contrast to a pure Transformer encoder, TransUNet adopts a hybrid CNN and Transformer architecture. The CNN backbone first performs hierarchical downsampling rather than using the ViT patches sized $P \times P$. Then the hidden feature is flattened into a sequence of linear projection through a $1 \times 1$ convolution. After patch embedding the embedded features are propagated to the Transformer layer. \cite{TransUnet} finds that the hybrid CNN-Transformer encoder performs better than using a pure Transformer. 

\subsection{Cascaded UNet-Structured Upsampler}
The decoder employs a cascaded upsampling architecture to decode hidden features from the linear vectors, followed by convolutional and sigmoid operations at the final stage. Through the upsampling process, the image dimensions transform from  $\frac{H}{2^N} \times \frac{W}{2^N}$ to $H \times W$. Each upsampling module consists of an upsampling operator, convolutional layers, and ReLU activation. 

Collectively, the synergistic integration of the encoder and decoder facilitates multi-resolution feature aggregation, forming a distinctive U-shaped architecture.

\addtolength{\topmargin}{0.05in}
\subsection{Loss Function Design}
In the construction of CKM, The electromagnetic wave intensity distribution in $\mathcal{M}$ manifests pronounced edge discontinuities induced by specular reflections and beamforming effects. To better capture these features using NNs, the proposed framework employs a designed composite loss function to guide the optimization process. As shown in (9), the total loss combines three distinct components with chosen weighting coefficients

\begin{equation}
	\label{eq:total_loss}
	\mathcal{L}_{\text{total}} = \underbrace{\lambda_1\mathcal{L}_{L2}}_{\text{pixel accuracy}} + \underbrace{\lambda_2\mathcal{L}_{\text{lap}}}_{\text{multi-scale detail}} + \underbrace{\lambda_3\mathcal{L}_{\text{edge}}}_{\text{structural preservation}},
\end{equation}
where $\{\lambda_1, \lambda_2, \lambda_3\} = \{1, 0.02, 0.01\}$ denote the relative importance weights for each component, determined through empirical validation.

\subsubsection{L2 (Mean Squared Error) Loss}
The foundation of our loss function is the standard L2 norm, which ensures basic pixel-level accuracy

\begin{equation}
	\label{eq:l2_loss}
	\mathcal{L}_{L2} = \frac{1}{N}\sum_{i=1}^{N}(y_{\text{pred}}^{(i)} - y_{\text{true}}^{(i)})^2,
\end{equation}
where $N = H \times W$ is the total number of pixels in the image of height $H$ and width $W$. This component provides a strong baseline for image reconstruction quality.

\subsubsection{Laplacian Pyramid Loss}
To preserve multi-scale structural fidelity, we adopt a 4-level Laplacian pyramid decomposition\cite{I2I}

\begin{equation}
	\label{eq:lap_loss}
	\mathcal{L}_{\text{lap}} = \frac{1}{C}\sum_{c=1}^{C}\sum_{l=1}^{4} w_l \| \mathcal{L}_l(y_{\text{pred}}^c) - \mathcal{L}_l(y_{\text{true}}^c) \|_1,
\end{equation}
where $C$ is the number of image channels,
 $w_l \in \{1.0, 0.5, 0.25, 0.125\}$ are level-dependent weights and
 $\mathcal{L}_l(\cdot)$ represents the Laplacian residual at level $l$.
The pyramid construction process involves: 1) downsampling via average pooling; and 2) upsampling with bilinear interpolation.
This multi-scale approach ensures that texture details are preserved across different spatial frequencies.

\subsubsection{Edge-Aware Loss}
For enhanced structural preservation, we derive a multi-scale edge-aware loss through Sobel gradient operators. First, compute horizontal ($G_x$) and vertical ($G_y$) gradients through convolution with Sobel kernels.
Then calculating edge strength maps with numerical stability constant $\epsilon = 10^{-6}$ yields
\begin{equation}
	E = \sqrt{G_x^2 + G_y^2 + \epsilon}.
\end{equation}
The final edge loss combines L1 differences across all channels is given by

\begin{equation}
	\label{eq:edge_loss}
	\mathcal{L}_{\text{edge}} = \frac{1}{C}\sum_{c}^{C} \|E_{\text{pred}}^c - E_{\text{true}}^c\|_1.
\end{equation}

\begin{figure*}[!htbp]
	\centering
	\begin{subfigure}[b]{0.19\textwidth}
		\includegraphics[width=\textwidth]{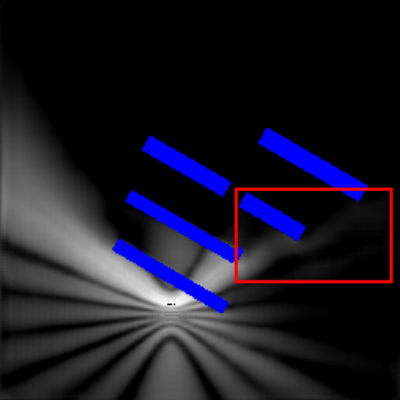}
		\caption{RME-GAN}
		\label{fig:rme-gan_a}
	\end{subfigure}
	\hfill
	\begin{subfigure}[b]{0.19\textwidth}
		\includegraphics[width=\textwidth]{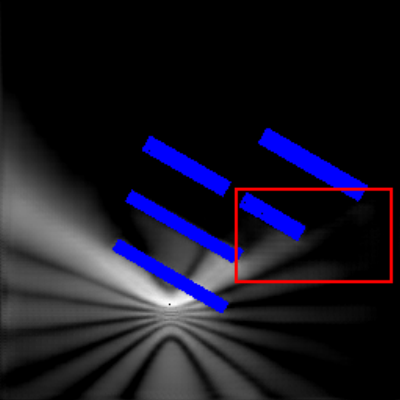}
		\caption{UNet}
		\label{fig:unet_b}
	\end{subfigure}
	\hfill
	\begin{subfigure}[b]{0.19\textwidth}
		\includegraphics[width=\textwidth]{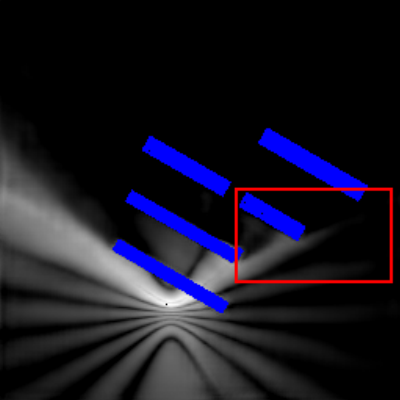}
		\caption{RadioWNet}
		\label{fig:radiownet_c}
	\end{subfigure}
	\hfill
	\begin{subfigure}[b]{0.19\textwidth}
		\includegraphics[width=\textwidth]{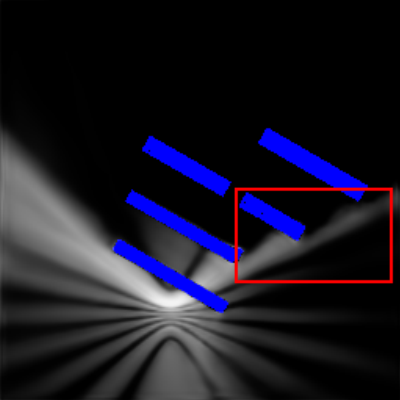}
		\caption{TransUNet}
		\label{fig:transunet_d}
	\end{subfigure}
	\hfill
	\begin{subfigure}[b]{0.19\textwidth}
		\includegraphics[width=\textwidth]{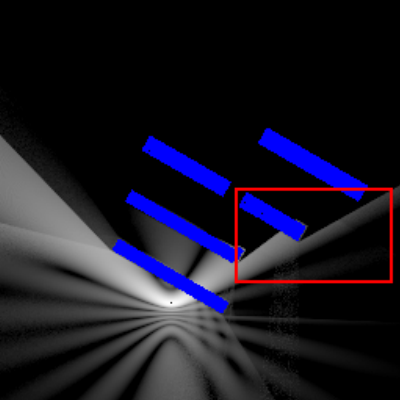}
		\caption{Ground truth}
		\label{fig:groundtruth_e}
	\end{subfigure}
	
	\vspace{0.5cm}
	
	\begin{subfigure}[b]{0.19\textwidth}
		\includegraphics[width=\textwidth]{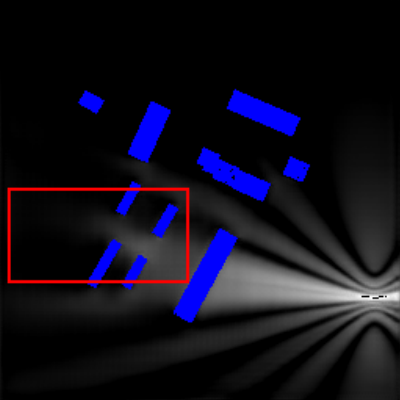}
		\caption{RME-GAN}
		\label{fig:rme-gan_f}
	\end{subfigure}
	\hfill
	\begin{subfigure}[b]{0.19\textwidth}
		\includegraphics[width=\textwidth]{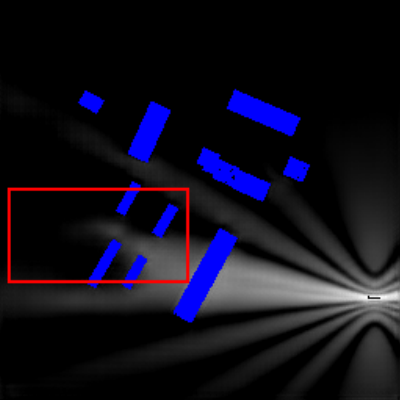}
		\caption{UNet}
		\label{fig:unet_g}
	\end{subfigure}
	\hfill
	\begin{subfigure}[b]{0.19\textwidth}
		\includegraphics[width=\textwidth]{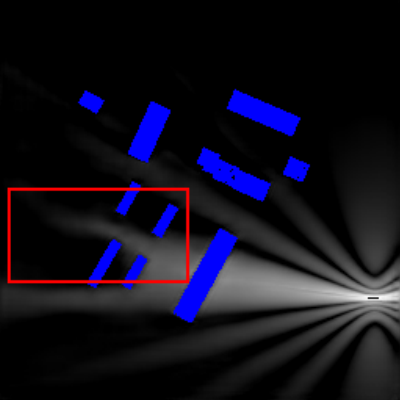}
		\caption{RadioWNet}
		\label{fig:radiownet_h}
	\end{subfigure}
	\hfill
	\begin{subfigure}[b]{0.19\textwidth}
		\includegraphics[width=\textwidth]{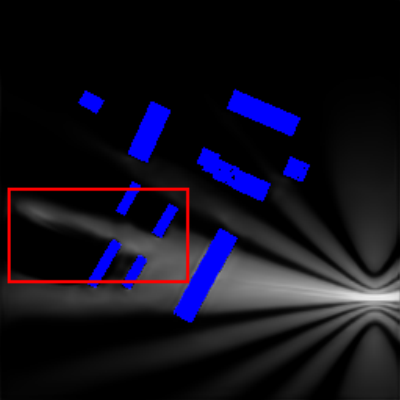}
		\caption{TransUNet}
		\label{fig:transunet_i}
	\end{subfigure}
	\hfill
	\begin{subfigure}[b]{0.19\textwidth}
		\includegraphics[width=\textwidth]{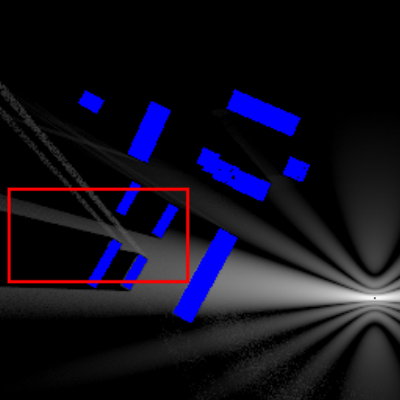}
		\caption{Ground truth}
		\label{fig:groundtruth_j}
	\end{subfigure}
	
	\caption{Visualization of predicted CKM and ground truth with specific beamforming.}
	\label{fig:radio_map_comparison}
\end{figure*}

\section{Simulations and discussion}
\subsection{Dataset and Simulation Setup}
The scene acquisition process leverages OpenStreetMap data\cite{Openstreetmap} as foundational input, from which we systematically extract 100 distinct urban topologies spanning diverse cities. Each geographical scene is rasterized into 256×256 pixel orthographic projections. These geospatial representations undergo electromagnetic property set in Blender through randomized material parameterization before integration into the Sionna ray-tracing environment \cite{Sionna}. 
For each static urban configuration, we implement uniform spatial sampling to generate 100 statistically independent BS deployments. Following BS placement, we employ Monte Carlo methods to distribute receiver nodes. The electromagnetic simulation incorporates $N$-dimensional DFT codebooks with all the altitudes fixed at $z = 1.5\text{m}$. Spatial pathloss aggregation is performed by computing the mean channel gain across all sampling points within each grid cell. Finally, the equivalent channel gain is normalized based on the maximum pathloss value and a dynamic range threshold $P_\text{thre}=40\text{dB}$, which can be expressed as
\begin{equation}
	P_{\text{norm}} = \frac{P_{\text{raw}} - P_{\text{thre}}}{P_{\text{max}} - P_{\text{thre}}}, 
\end{equation}
where the raw value $P_{\text{raw}} < P_{\text{thre}}$ will be set to zeros. Buildings and the BS are assigned a value of 1 to indicate their locations. 

Following established methodologies in SOTA approaches, we construct the network input by concatenating environmental data with BS location to form a 2×256×256 dimensional map. The simulation configuration assigns $N=8$ antennas to each BS. These inputs are then processed through TransUNet architecture, which combines a ResNet-based CNN feature extractor (with 2 skip connections) with a 12-layer Transformer module, as detailed in Fig. 3. For experimental hyperparameters, we partition the datasets into training, validation, and test subsets with 8:1:1 ratio. All models are trained using the Adam optimizer for 100 epochs with an initial learning rate of 0.0001, which decays by a factor of 0.2 every 20 epochs, and a batch size of 16 is maintained throughout the training process.

\begin{figure}[!t]
	\centering
	\begin{subfigure}[b]{1\columnwidth}  
		\centering
		\includegraphics[width=\linewidth]{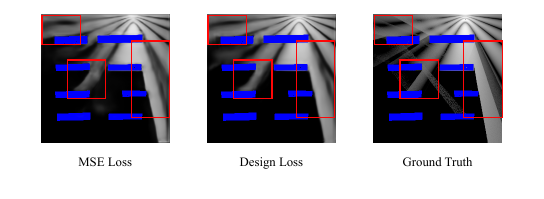}
		\label{fig5a}
	\end{subfigure}
	\hfill
	\begin{subfigure}[b]{0.9\columnwidth}
		\centering
		\includegraphics[width=\linewidth]{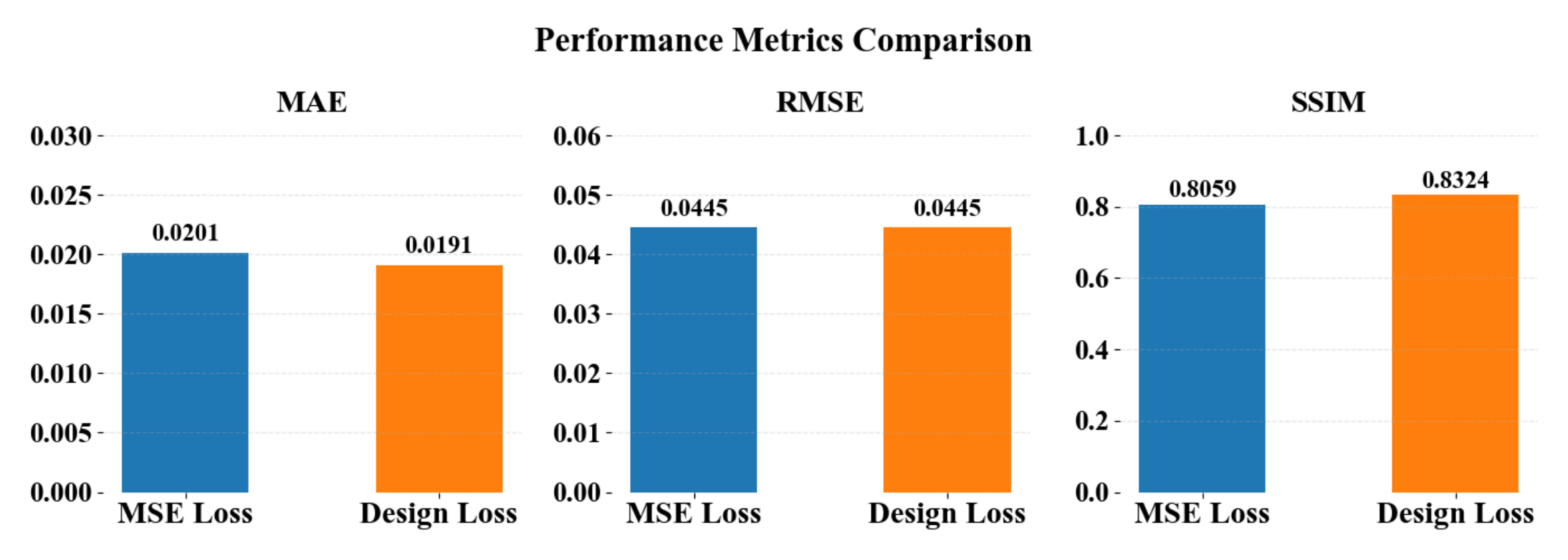}
		\label{fig5b}
	\end{subfigure}
	
	\caption{Overall comparison between different loss functions.}
	\label{fig5}
\end{figure}

\begin{table}[h]
	\centering
	\caption{Comparison of different methods} \label{six}
	\setlength{\tabcolsep}{4pt}
	\renewcommand{\arraystretch}{1.5}
	\begin{tabular}{l @{\hspace{1em}} *{5}{c}}
		\hline
		\multicolumn{1}{l}{Methods} & RME-GAN & UNet & RadioWNet & TransUNet \\
		\hline
		MAE      & 0.0268 & 0.0243 & 0.0220  & $\textbf{0.0191}$ \\
		NMSE     & 0.0965 & 0.0883 & 0.0824  & $\textbf{0.0570}$ \\
		RMSE     & 0.05910 & 0.0561 & 0.0537 & $\textbf{0.0445}$ \\
		PSNR$\uparrow$ & 24.7859 & 25.2809 & 25.7044 & $\textbf{27.3501}$ \\
		SSIM$\uparrow$ & 0.7823	 &	0.8023 & 0.8158	&  $\textbf{0.8324}$	\\
		\hline
	\end{tabular}
\end{table}

\subsection{Discussion}
To evaluate the performance of the proposed 
TransUNet model, we conduct
comprehensive comparisons with SOTA methods. The comparative methods are implemented with the same parameter setting, including UNet, RadioWNet\cite{RadioUnet}, CKM-GAN\cite{GAN}. We deliberately exclude the additional observational inputs used in the original paper to ensure a fair comparison under identical input conditions. All frameworks are employed using identical hyperparameters and training protocols. The evaluation metrics include RMSE, NMSE, MAE, PSNR, and SSIM \cite{I2I}.

Experimental results in Table I demonstrate that TransUNet consistently outperforms baseline methods across all evaluation metrics. Specifically, it exhibits a 17\% improvement in RMSE compared to RadioWNet. Furthermore, Fig. 4 demonstrates TransUNet's enhanced capability in resolving fine-grained details within critical regions, successfully predicting subtle features that elude comparative methods. As evidenced by Fig. 5, the proposed loss function demonstrates robust performance to pure MSE loss in beam edge regions, which achieves simultaneous MAE reduction and SSIM enhancement while maintaining RMSE.

Moreover, experimental verification shows that the TransUNet model achieves an inference time of 0.017 seconds per prediction. This represents a fivefold reduction compared to the 0.098 seconds required by accelerated ray-tracing algorithms in Sionna. The proposed approach eliminates the need for pilot signals to estimate pathloss, thereby improving spectral utilization efficiency.

In conclusion, the proposed architecture not only demonstrates measurable improvements in quantitative metrics but also achieves higher fidelity in reproducing the physical characteristics of electromagnetic wave propagation phenomena.
The experimental results successfully demonstrate distinct coverage patterns across different beams, confirming the neural network's capability to construct CKMs for diverse beamforming vectors. Leveraging the NNs' rapid inference capability, TransUNet enables efficient CKM prediction, which facilitates beamforming design validation while significantly reducing the required beam scanning range. This approach consequently leads to substantial reductions in communication overhead.

\section{Conclusion}
This paper proposed a novel TransUNet-based framework for beamforming-aware CKM construction under diverse DFT codebook configurations. Simulations demonstrated substantial performance gains over SOTA DL methods, as evidenced by significant improvements in RMSE and other key metrics. The study made two key contributions: 1) it established the feasibility of NN-based CKM prediction in multi-antenna beamforming systems, and 2) it demonstrated TransUNet's superior performance in electromagnetic environment modeling. Future research was suggested to explore CKM-assisted beamforming optimization and beam training overhead reduction through predictive CKM utilization.

\balance

\bibliographystyle{IEEEtran}
\bibliography{reference.bib}

\end{document}